# Electrical transport in epitaxially grown undoped and Si-doped degenerate GaN films

Mohammad Monish[*] and S. S. Major[†]

*Department of Physics, Indian Institute of Technology Bombay, Mumbai 400076, India*

*Abstract*

Undoped and Si-doped GaN films were grown epitaxially on sapphire by reactive rf sputtering of GaAs (and Si) in Ar-$N_2$ mixture. The resistivity of undoped GaN film grown at 100% $N_2$ was ~$2 \times 10^5$ $\Omega$ cm, which reduced to ~1 $\Omega$ cm in Si-doped film, revealing the effect of Si doping. With decrease of $N_2$ from 100% to 75%, the carrier concentration of Si-doped films increased from ~$7 \times 10^{18}$ cm$^{-3}$ to ~$2 \times 10^{19}$ cm$^{-3}$, but remained practically unchanged as $N_2$ was decreased to 20%, which is explained by effects due to saturation of Si doping and increase of Ga interstitials as well as compensation by N interstitials and Ga vacancies. Undoped and Si-doped films grown below 20% $N_2$ displayed similar carrier concentrations (~$10^{20}$ cm$^{-3}$), due to dominance of Ga interstitials. Both undoped and Si-doped films were degenerate and displayed increase of mobility with carrier concentration and temperature, which was analyzed by the combined effect of ionized impurity and dislocation scattering, using compensation ratio as fitting parameter. At carrier concentrations $\lesssim 10^{19}$ cm$^{-3}$, the mobility was governed by both ionized impurity and dislocation scattering, while at higher carrier concentrations, ionized impurity scattering was found to dominate, limited by compensation due to acceptors. In spite of the degenerate character, the films displayed a small decrease of carrier concentration with temperature, along with a nearly linear decrease of mobility, which are explained by a marginal increase of compensation ratio with decrease of temperature, taking into account the band edge fluctuation effects.

---

[*] monish.iitb@gmail.com
[†] Author to whom any correspondence should be addressed: syed@iitb.ac.in
Tel.: +91-22-25767567; Fax: +91-22-25767552

# 1. Introduction

Following the rapid advancement of GaN based devices, ranging from short wavelength optoelectronics and white light sources to high power and high frequency transistors [1-4], the progress of GaN as the next generation semiconductor has been constantly driving improvements in the processing of this versatile material. The ever-increasing scope of GaN calls for exploring non-conventional approaches of material growth and controlled in-situ doping, as well as the improved understanding of its electronic properties, particularly those of the films grown by alternative methods. Epitaxial GaN films are traditionally grown by molecular beam epitaxy (MBE), hydride vapour phase epitaxy (HVPE) and metal-organic chemical vapor deposition (MOCVD), in spite of the large lattice and thermal expansion coefficient mismatch between GaN and sapphire [5]. The hetero-epitaxial growth of GaN on sapphire leads to high density of threading dislocations, in the range of $10^8$ - $10^{10}$ cm$^{-2}$ [6-8], but unlike most other compound semiconductors, such GaN epilayers have been used to fabricate high quality optoelectronic devices [9,10]. Although the density of threading dislocations has often been limited to the range of $10^6$ - $10^7$ cm$^{-2}$ in MOCVD [11] and MBE [12] grown GaN films, these techniques have some limitations, such as, high temperature operation in MOCVD and the scalability and high cost of MBE. Hence, alternative methods, such as sputtering remain attractive and continue to be explored for the growth of GaN [13-17]. In recent years, sputtering has demonstrated considerable promise for the growth of epitaxial GaN films, although with threading dislocations in the range of $10^9$ - $10^{12}$ cm$^{-2}$ [18-20]. It has been employed for the growth of both undoped [18,21] and doped (with Si, Ge and Mg) [19,22-24] GaN films, as well as for fabrication of GaN based LEDs [25]. Pulsed sputtering [19,22,24] has been used with a base pressure of ~$5 \times 10^{-10}$ Torr to homoepitaxially grow doped GaN films on MOCVD/HVPE grown Fe-doped GaN templates, which display mobility values in the range of 5 - 1000 cm$^2$V$^{-1}$s$^{-1}$. Si has been the most common *n*-type dopant in GaN, which has yielded films with carrier concentrations as high as ~$3 \times 10^{20}$ cm$^{-3}$ by different techniques [22,26,27] and is often used to reduce the parasitic contact resistance of various device structures [28,29]. Unintentionally doped GaN epitaxial films grown by sputtering have also been recently reported [21] to display high carrier concentrations of ~$10^{20}$ cm$^{-3}$.

The mobility of MOCVD and MBE grown non-degenerate GaN films possessing threading edge dislocations $\lesssim 10^8$ cm$^{-2}$ is usually reported to be in the range of $10^2$ - $10^3$ cm$^2$V$^{-1}$s$^{-1}$, and is predominantly controlled by the lattice and ionized impurity scattering mechanisms [30-32] Look *et al*. [30] analyzed the temperature dependent mobility and carrier concentration of high quality MOCVD grown *n*-type GaN films with room temperature mobility of ~600 cm$^2$V$^{-1}$s$^{-1}$, by using a two-donor/one-acceptor model and considering scattering mechanisms due to polar optical-mode, acoustic-mode deformation-potential, piezoelectric-potential as well as ionized impurities/defects. They concluded that mobility analysis is essential for accurately determining the acceptor concentration or compensation ratio. Tang *et al*. [32] also fitted the mobility vs. temperature data of MBE grown Si-doped GaN films with room temperature mobility of ~200 cm$^2$V$^{-1}$s$^{-1}$ and different acceptor concentrations by considering lattice and ionized impurity scattering mechanisms. They showed that the low temperature mobility was mainly limited by ionized impurity scattering and was sensitive to the presence of compensating acceptors. However, near room temperature, the mobility was found to follow an empirical dependence on temperature (~T$^{-1.5}$), which was attributed to crystalline defects, such as dislocations, domain boundaries and strain induced fields [32]. As mentioned above, although hetero-epitaxial GaN films on sapphire with threading dislocation density approaching 10$^{10}$ cm$^{-2}$ have been successfully used to fabricate optoelectronic devices [9], the high density of dislocations has been shown to significantly affect their mobility, bringing it down to the range of 100 - 200 cm$^2$V$^{-1}$s$^{-1}$ [33-35]. It is also known that the dangling bonds associated with Ga vacancies [36] and V$_{Ga}$ – O$_N$ complexes [37] introduce acceptor centers and result in the formation of charged edge dislocation lines in *n*-type GaN. Weimann *et al*. [33] have shown that the influence of charged dislocations on the scattering of electrons becomes significant when the dislocation density is ≳10$^9$ cm$^{-2}$. They modelled the low mobility of MBE grown Si-doped GaN films having different dislocation densities by considering the statistics of trap occupancy at different doping levels. Weimann *et al*. [33] and Ng *et al*. [34] have also shown that the mobility of MBE grown Si-doped GaN films increased with increase of carrier concentration. They explained this behaviour by using a relation, initially derived by Podor [38] for *n*-type Ge, which considers that the Coulomb scattering of carriers by charged dislocations is screened by conduction electrons and possibly charged donors [33,34]. According to their results, the mobility of Si-doped GaN films having different edge dislocation densities in the range of 10$^9$ - 10$^{11}$ cm$^{-2}$ followed a family of bell-shaped curves, wherein it increased with increasing carrier concentration up to values as high as ~10$^{20}$ cm$^{-3}$ (depending on the dislocation density), beyond which it decreased due to the effect of ionized impurities. Subsequently, Look and Sizelove [35] developed a quantitative model, which includes scattering potentials due to optical and acoustic modes as well as screened Coulomb potential due to charged dislocations and ionized impurities. They concluded that threading edge dislocations in GaN are electrically active and used their model to fit the temperature dependent

Hall effect data of MBE grown GaN films with low ($4 \times 10^8$ cm$^{-2}$) and high ($2 \times 10^{10}$ cm$^{-2}$) dislocation densities.

We have recently demonstrated the epitaxial growth of highly conducting, unintentionally doped GaN [21] films by reactive sputtering of GaAs in Ar-N$_2$ sputtering atmosphere. It was shown that the resistivity of undoped GaN films depends strongly on nitrogen partial pressure, decreasing drastically from ~$2 \times 10^5$ Ω cm to ~$2 \times 10^{-3}$ Ω cm with decrease of N$_2$ from 100% to 10% in sputtering atmosphere. Subsequent studies [20,39] have revealed that the high resistivity of GaN films grown at ~100% N$_2$ was due to the compensation by N interstitial acceptors, while the drastic decrease in resistivity of the films grown at low N$_2$ percentages (~10%) was caused by the substantial decrease of N interstitials, along with the formation of Ga interstitials, which act as donors. Another work from our group on co-sputtered Si-doped GaN films [23] has also shown that their resistivity varies in the range of $10^{-3}$ - 1 Ω cm, depending on the Si area coverage of GaAs target and N$_2$ percentage in sputtering atmosphere. These results have shown that both undoped and Si-doped sputtered GaN films are degenerate, having high carrier concentrations (>$10^{18}$ cm$^{-3}$) with modest mobility values [33,40]. Our earlier studies [20,41] on the microstructure of undoped and Si-doped GaN films grown epitaxially by sputtering have also shown that these films possess a high density of edge dislocations in the range of $10^{11}$ - $10^{12}$ cm$^{-2}$. It may be mentioned here, that in spite of the recent progresses in the growth of epitaxial GaN films by sputtering and promising device applications [25], studies related to their electronic properties and dependence on microstructure and defects, remain rather limited, compared to the enormous scientific literature that exist on MBE and MOCVD grown GaN. The present work attempts a comparative study of the electrical properties of undoped and Si-doped, *n*-type degenerate GaN films grown epitaxially by reactive sputtering at different partial pressures of N$_2$, and in particular, investigates the concomitant increase of room temperature mobility with carrier concentration, observed in both cases. Furthermore, temperature dependent measurements of resistivity, carrier concentration and mobility were carried out to investigate the mechanism of electrical transport in these films. A systematic analysis of the dependence of mobility on carrier concentration and temperature has been carried out by considering the combined effects of scattering due to ionized impurities and charged edge dislocation lines. These aspects of the electrical properties of undoped and Si-doped GaN films grown epitaxially by sputtering have not been much explored, and in general, not for *n*-type, degenerate GaN films, possessing high density of dislocations and point defects. Our results show that the mobility behaviour of highly degenerate GaN films is governed by the relative contributions to scattering due to ionized impurities and charged edge dislocations, where the former is significantly influenced by changes in compensation ratio.

## 2. Experimental details

Undoped and Si-doped epitaxial GaN films were grown by reactive rf magnetron sputtering of a GaAs target, as described earlier [20,23]. For the growth of Si-doped GaN films, the Si area coverage of the erosion track of GaAs was fixed at ~5%, based on earlier results, which have indicated the deterioration of the structural quality of films grown with larger area coverages [23]. Both undoped and Si-doped GaN films were grown at 700 °C in Ar-$N_2$ atmosphere (total pressure, 0.86 Pa) over approximately 50 nm thick GaN buffer layers on $c$-sapphire substrate, which were grown in 100% $N_2$ at 300 °C. The films were grown at different nitrogen partial pressures in the range of 0.06 Pa to 0.86 Pa, by changing the corresponding $N_2$ percentage in sputtering atmosphere from 7% to 100%. Accordingly, the $N_2$ percentage during growth will be used to label the films grown at various partial pressures of $N_2$ in this work. The RF power was adjusted to maintain the growth rate of films at $(1.0 \pm 0.1)$ μm hr$^{-1}$, as the Ar percentage was increased. The thickness of all the films was in the range of 850 - 900 nm.

Lake Shore 8400 series Hall measurement system was used to measure the resistivity and Hall coefficient (at a magnetic field of 0.64 Tesla) in van der Pauw's geometry at room temperature and in the temperature range of 10 - 300 K. The carrier concentration and Hall mobility of the films were obtained from these measurements, performed in the AC field Hall measurement mode, in which, the instrument is capable of measuring high carrier concentration up to ~$10^{23}$ cm$^{-3}$ and Hall mobility as low as ~$10^{-3}$ cm$^2$V$^{-1}$s$^{-1}$. X-ray photoelectron spectroscopy (XPS) measurements were performed with a Kratos Analytical Axis Supra ESCA instrument, using a monochromatic Al K$_\alpha$ (1486.6 eV) source. Charge neutralization was employed during the measurement and the binding energy scale was referenced to C 1$s$ peak at 284.8 eV. ESCApe software was used for the analysis of core level spectra, which were fitted with Shirley-type background and a combination of Gaussian and Lorentzian line shape components for the peak envelope. High resolution x-ray diffraction (HRXRD) measurements were carried out with a Cu rotating anode (9 kW), RIGAKU-Smart Lab-X-Ray diffractometer, equipped with a 2-bounce Ge (220) monochromator, to calculate the density of edge dislocations from mosaic twist, obtained by in-plane measurements of $(11\bar{2}0)$ reflection, as described elsewhere [20,41].

## 3. Results and discussion

The structural data of undoped [20] and Si-doped [41] GaN films grown at different $N_2$ percentages in sputtering atmosphere have been reported earlier. In both cases, the out-of-plane $\omega$-$2\theta$ high-resolution scans have shown only (000$l$) reflections of GaN, confirming the formation of single wurtzite phase of GaN, with nearly complete $c$-axis orientation. The phi ($\phi$) scans of (10$\bar{1}$1) reflection in both cases [20,41] have confirmed the epitaxial character of the films, conforming to the in-plane relationship of GaN [11$\bar{2}$0]∥α-Al$_2$O$_3$[10$\bar{1}$0] [42] and alignment with the oxygen sub-lattice of sapphire. XPS measurements of undoped GaN films [39] have shown that the N/Ga ratio decreased substantially with decrease of $N_2$ percentage in sputtering atmosphere. Furthermore, XPS and secondary ion mass spectroscopy (SIMS) measurements of sputtered undoped GaN films [20,39] revealed the absence of arsenic but the presence of oxygen impurity in the films. In the case of Si-doped GaN films, EDX measurements have shown the presence of ~2 at.% Si in all the films along with oxygen impurity (~3 at.%) and confirmed the absence of arsenic [41]. It was also found that the N/Ga ratio in Si-doped films decreased with decrease of $N_2$ percentage in sputtering atmosphere, similar to the observations made for undoped GaN films [39,41].

Hall measurements were carried out to obtain the resistivity, mobility and carrier concentration of undoped and Si-doped GaN films grown at different $N_2$ percentages in sputtering atmosphere, and the results are shown in figures 1-3. Figure 1 shows the variation of the resistivity of undoped and Si-doped GaN films grown in the range of 7% - 100% $N_2$. Undoped GaN film grown at 100% $N_2$ displays a large resistivity of $\sim 2 \times 10^5$ Ω cm, which decreases monotonously and drastically with decrease of $N_2$ percentage to attain a low value of $\sim 2 \times 10^{-3}$ Ω cm for the films grown in the range of 8% - 10% $N_2$, as reported earlier [21]. In contrast, the Si-doped GaN film grown at 100% $N_2$ shows a resistivity of ~1 Ω cm, which is five orders of magnitude less than that of the undoped GaN film, grown under similar conditions. The resistivity of the Si-doped GaN films grown at ~75% $N_2$ decreases substantially to $\sim 6 \times 10^{-2}$ Ω cm, but remains practically unchanged with further decrease of $N_2$ to ~20%. Interestingly, the resistivities of Si-doped GaN films grown at 20% or lower $N_2$ are close to those of the undoped GaN films and attain nearly equal values ($\sim 2 \times 10^{-3}$ Ω cm) in the films grown at ~10% $N_2$. It is also noticed that the resistivities of both undoped and Si-doped GaN films grown at lower $N_2$ percentages tend to increase and become substantially larger ($\sim 2 \times 10^{-2}$ Ω cm) in the case of the films grown at 7% $N_2$.

The carrier concentration of undoped and Si-doped GaN films grown at different $N_2$ percentages are shown in figure 2. It may be mentioned that undoped GaN films grown at 50% or higher $N_2$ displayed high resistivity, which made it difficult to perform Hall measurements, hence, the carrier concentration and mobility were measured only for the films grown at ≲40% $N_2$. On the other hand, figure 2 shows that the carrier concentration of Si-doped GaN film grown at 100% $N_2$ is ~$7 \times 10^{18}$ cm$^{-3}$, which increases to ~$2 \times 10^{19}$ cm$^{-3}$ for the film grown at 75% $N_2$. However, with further decrease of $N_2$ to ~20%, the carrier concentration remains practically unchanged, displaying a plateau-like behaviour, similar to that seen in the case of resistivity. Interestingly, with decrease of $N_2$ below 20%, the carrier concentrations of both undoped and Si-doped GaN films increase, displaying comparable values, consistent with their resistivity behaviour and attain the highest carrier concentration of ~$3 \times 10^{20}$ cm$^{-3}$, for the films grown at 7% $N_2$.

The above results show that firstly, the Si-doped GaN films grown with 20% - 100% $N_2$ possess substantially higher carrier concentrations compared to the corresponding undoped GaN films. This is particularly evident for the films grown in the range of 20% - 40% $N_2$ (figure 2), and is clearly the effect of substitutional doping of GaN with Si. Secondly, the increase of carrier concentration of Si-doped GaN films with the initial reduction of $N_2$ percentage and the plateau-like behaviour seen in the range of 20% - 75% $N_2$ may be explained as follows. It has earlier been shown [20,41] that both undoped and Si-doped GaN films grown at 100% $N_2$ possess large hydrostatic strain due to the presence of N interstitials (known to be acceptors in GaN [43,44]), which decreased with decrease of $N_2$ percentage in sputtering atmosphere. The high resistivity of undoped GaN films grown at 100% $N_2$ and its decrease with $N_2$ percentage was hence primarily attributed to the strong compensation by N interstitials and their subsequent decrease [39]. Thus, the low carrier concentration of Si-doped GaN films grown at 100% $N_2$ is also attributed to the compensation by N interstitials and its subsequent increase in the film grown at 75% $N_2$ to the substantial decrease of N interstitials, as shown earlier [41]. As mentioned above, the Si content of all the doped films grown at different $N_2$ percentages is nearly the same [41] and hence, the plateauing of carrier concentration at ~$2 \times 10^{19}$ cm$^{-3}$ primarily implies the saturation of Si doping at a low efficiency of ~2%. However, in order to further understand the plateau-like behaviour of the carrier concentration of Si-doped GaN films and its increase at lower $N_2$ percentages, XPS measurements of high resolution Ga 3$d$ core level spectra for typical Si-doped GaN films grown at 100% and 10% $N_2$ were carried out, which are shown in figure 4. The spectra were deconvoluted and fitted into three main component peaks in the range of (20.5 - 20.7 eV), (19.5 - 19.7 eV) and (18.2 - 18.4 eV), attributed respectively to Ga-O bonding [45,46], Ga-N bonding [45,46] and Ga-Ga (metallic/uncoordinated Ga) [47], along with a small peak due to N 2$s$ [48] in the range of 16.2 - 16.4 eV. It is found that with decrease of $N_2$ from 100% to 10%, the uncoordinated/interstitial Ga to Ga-N ratio increases from ~0.1 to ~0.3, similar to the corresponding values reported earlier [39] for undoped GaN films. It is also noteworthy that our x-ray absorption spectroscopy (XAS) and XPS measurements [39] of undoped GaN films have shown that both Ga interstitials and Ga vacancies increase with decrease of $N_2$ percentage and, Ga interstitials, in particular, become prominent in the films grown at low $N_2$. Based on these observations, it was inferred [39] that Ga interstitials were primarily responsible for the *n*-type doping of undoped GaN films, while acceptors due to Ga vacancies [49] also contributed to compensation. In the case of Si-doped films, considering Si activation to have practically reached its limit, with decrease of $N_2$ percentage in the plateau region, there is an increasing contribution of Ga interstitials as donors. On the other hand, presuming that, while N interstitials decrease with $N_2$ percentage, there is an increase of Ga vacancies, which also contribute to compensation, thus resulting in the plateau-like behaviour. However, the substantial and nearly similar increase of carrier concentration of both undoped and Si-doped GaN films grown below 20% $N_2$, further indicates that a different doping mechanism dominates in the Si-doped GaN films grown at lower $N_2$ percentages, which is the same as in the case of undoped GaN films. It is thus inferred, that the high carrier concentration (~$10^{20}$ cm$^{-3}$) in both undoped and Si doped GaN films grown below 20% $N_2$, is caused primarily by the substantial increase of Ga interstitials, which act as the dominant donor species in both cases.

Figure 3 shows the variation of room temperature Hall mobility of undoped and Si-doped GaN films grown at different $N_2$ percentages in sputtering atmosphere. As in the case of carrier concentration, the mobility of undoped GaN films could be measured only for those grown in the range of 7% - 40% $N_2$, while it was measurable for all the Si-doped films. The Si-doped GaN film grown at 100% $N_2$ shows a low mobility of ~2 cm$^2$V$^{-1}$s$^{-1}$, which increases to ~10 cm$^2$V$^{-1}$s$^{-1}$ for the films grown in the range of 30% - 50% $N_2$. In comparison, undoped GaN films grown in the range of 30% - 40% $N_2$ display substantially lower mobility than those of Si-doped GaN films grown under similar conditions. Interestingly, at ~20% and lower $N_2$, the mobility of both undoped and Si-doped GaN films displays nearly equal values and increases thereafter with decrease of $N_2$ percentage, finally attaining values in the range of 23 - 25 cm$^2$V$^{-1}$s$^{-1}$ for the films grown at 10% $N_2$. However, as $N_2$ is further decreased below 10%, both undoped and Si-doped GaN films show a drastic decrease of mobility to ~1 cm$^2$V$^{-1}$s$^{-1}$ for the films grown at 7% $N_2$, consistent with the observed increase in resistivity of the films. The substantial decrease of mobility is attributed to the possible formation of GaAs$_x$N$_{1-x}$ in the films grown at low N$_2$ percentages, as reported earlier [50]. It may be mentioned that our microstructural and Raman studies reported recently [20], have indicated that GaN films grown at low N$_2$ percentages (~10%) display substantially enhanced disorder and poorer microstructural quality.

The results presented in figures 1-3 thus reveal a strong dependence of the electrical parameters of both undoped and Si-doped GaN films on the N$_2$ percentage in sputtering atmosphere and more importantly, display an increase of mobility with carrier concentration over a large range of $10^{18}$ - $10^{20}$ cm$^{-3}$. As discussed in the Introduction, such an increase of mobility with carrier concentration has been attributed mainly to the scattering of electrons due to charged edge dislocations in epitaxial GaN films. However, the concurrent increase of carrier concentration and mobility, seen in both undoped and Si-doped GaN films grown by sputtering in the present study is particularly interesting for two reasons. Firstly, the critical carrier concentration at which semiconductor-metal transition (based on Mott's criterion [51,52]) occurs in GaN is estimated to be in the range of $10^{18}$ - $10^{19}$ cm$^{-3}$, implying that both undoped and Si-doped GaN films, which display carrier concentration $\gtrsim 2 \times 10^{18}$ cm$^{-3}$ can be treated as degenerate [9,31]. Secondly, both undoped and Si-doped films also possess relatively large density of edge dislocations ($10^{11}$ - $10^{12}$ cm$^{-2}$). It may also be mentioned that the high carrier concentration of both undoped and Si-doped GaN films implies the substantial presence of ionized impurities due to charged donors and acceptors, which may also be responsible for the formation of charged dislocation lines [36]. Hence, it is likely that the mobility of undoped and Si-doped GaN films is determined by the presence of both ionized impurities as well as charged edge dislocations.

In semiconductor films with such high carrier concentrations, the scattering due to ionized impurities is likely to be governed by the degenerate form of Brooks-Herring relation [53], which is not explicitly temperature dependent, but depends significantly on the compensation due to acceptor centers. The contribution of ionized impurity scattering to mobility can accordingly be estimated by using the degenerate form of Brooks-Herring relation, as modified by Look *et al*. [53,54]

$$\mu_{ii} = \frac{24\pi^3\varepsilon^2\hbar^3 n}{Z^2 m^{*2} e^3 N_{ii}[\ln(1+\xi)-\xi/(\xi+1)]} \quad (1)$$

where, Z is the ionization charge (taken as 1 for donors and acceptors [54]) and $\xi = (3^{1/3}4\pi^{8/3}\varepsilon\hbar^2 n^{1/3})/(e^2 m^*)$. $N_{ii}$ ($= N_D^+ + N_A^- \simeq 2N_A^- + n$) is the total concentration of ionized charge centers, where $N_D^+$ and $N_A^-$ are the donor and acceptor concentrations, respectively and $n$ is the measured carrier concentration. Furthermore, a fitting parameter, $r\ (=N_A^-/N_D^+)$ is defined as the compensation ratio, which implies that in equation 1, $n/N_{ii}$ can be taken as $(1 - r)/(1 + r)$ [54]. As mentioned above, sputtered GaN films possess a high density of acceptor centres, which implies that the value of $r$ in these films may attain significantly large values.

The increase of mobility with carrier concentration has generally been attributed to scattering of electrons by charged edge dislocations [33,34], as originally suggested by Podor [38] for *n*-type Ge, considering that the Coulomb scattering of carriers by charged dislocations is screened by conduction electrons and possibly charged donors [33,34]. Subsequently, the dislocation scattering controlled mobility in degenerate systems has been modified by Look *et al*. [40] as

$$\mu_{dis} = \frac{4 \times 3^{2/3} ec^2 n^{2/3}}{\pi^{8/3} N_{dis} \hbar} \left[ 1 + \frac{2 \times 3^{1/3} \pi^{8/3} \hbar^2 \varepsilon n^{1/3}}{e^2 m^*} \right]^{3/2} \quad (2)$$

where, $N_{dis}$ is the density of edge dislocations, $c$ is the lattice constant (here, $c$ is taken to be equal to the lattice parameter along *c*-axis of GaN), $m^*$ is the electron effective mass and $\varepsilon$ is the static dielectric constant.

Thus, in order to analyze the dependence of the mobility of undoped and Si-doped GaN films on carrier concentration, considering the combined effect of ionized impurity and dislocation scattering mechanisms, the density of edge dislocations (dominant in hexagonal systems [55], including GaN [56]) in all the films was obtained by high resolution in-plane $\phi$-rocking curves of $(11\bar{2}0)$ reflection, using the methodology reported earlier [20,41]. Figure 5 shows the density of edge dislocations for both undoped and Si-doped GaN films grown in the range of 10% - 100% $N_2$. It is noted that both undoped and Si-doped GaN films display nearly similar dependences of the density of edge dislocations on $N_2$ percentage. The undoped and Si-doped GaN films grown at 30% - 100% $N_2$ display high density of edge dislocations $(1 - 3) \times 10^{12}$ cm$^{-2}$, which decreases substantially with decrease of $N_2$ percentage, attaining values of ~4 $\times 10^{11}$ cm$^{-2}$ for the films grown at 10% $N_2$, as also reported earlier [20,41].

Following the above, the dependence of the mobility of undoped and Si-doped GaN films on carrier concentration has been analyzed by considering the combined effect of ionized impurity and dislocation scattering mechanisms in degenerate semiconductors and the results are presented in figure 6. The variation of the measured mobility with carrier concentration for undoped and Si-doped GaN films grown at different $N_2$ percentages are shown in figure 6(a) and (b), respectively. It may be noted that several films were grown at each $N_2$ percentage to verify the reproducibility of data, the variation of which, is indicated by appropriate error bars.

The dependence of mobility on carrier concentration is found to be nearly similar for undoped and Si-doped GaN films, changing from ~2 cm$^2$V$^{-1}$s$^{-1}$ to ~25 cm$^2$V$^{-1}$s$^{-1}$, as the carrier concentration changes over two orders of magnitude from ~10$^{18}$ cm$^{-3}$ to ~10$^{20}$ cm$^{-3}$. In the following analysis, the values of $\mu_{ii}$ for undoped and Si-doped GaN films, were calculated from equation 1 and the corresponding values of $\mu_{dis}$ were calculated from equation 2. The compensation ratio '$r$' was used as a fitting parameter in equation 1, to arrive at the measured mobility ($\mu$) by invoking Matthiessen's rule ($\mu^{-1} = \mu_{ii}^{-1} + \mu_{dis}^{-1}$). The calculated values of $\mu_{dis}$ are shown in figures 6(a) and 6(b), which were obtained by using the average values of $N_{dis}$ (from figure 5) and the average values of carrier concentration of the films grown at different N$_2$ percentages. For the calculation of $\mu_{dis}$, the values of $\varepsilon$ and $m^*$ for GaN were taken as 8.9$\varepsilon_o$ and 0.20$m_0$, respectively [57]. The corresponding values of $\mu_{ii}$ were calculated from equation 1 and are also shown in figures 6(a) and 6(b). The values of $r$ used for fitting the mobility data are plotted against carrier concentration, and shown as insets of figures 6(a) and 6(b). It is seen from figure 6 that the measured mobility values of both undoped and Si-doped GaN films are smaller than the calculated values of $\mu_{dis}$ and $\mu_{ii}$. In the low carrier concentration range ($\lesssim$10$^{19}$ cm$^{-3}$), the values of $\mu_{dis}$ and $\mu_{ii}$ are nearly comparable, for both undoped and Si-doped GaN films. Hence, it is inferred that the low values of measured mobility in this range are determined by the combined effect of ionized impurity scattering and dislocation scattering. However, in the higher carrier concentration range (10$^{19}$ - 10$^{20}$ cm$^{-3}$), $\mu_{dis}$ increases substantially, which may be partly due to the decrease of $N_{dis}$, with decrease of N$_2$ percentage, as shown in figure 5. In these cases, although $\mu_{ii}$ also increases due to the decrease of compensation ratio, but it remains much smaller than the corresponding values of $\mu_{dis}$. It is thus inferred that at high carrier concentrations, $\mu_{ii}$ predominantly determines the overall mobility of the films, which is not significantly affected by dislocation scattering. This behaviour is in agreement with the results of Weimann *et al.* [33] for MBE grown Si-doped GaN films, who have also shown that at low carrier concentrations, the mobility is limited by scattering at charged dislocations, while ionized impurity scattering dominates at higher carrier concentrations. Figure 6 also shows that the values of compensation ratio are quite high (~0.9) for both undoped and Si-doped GaN films having carrier concentrations $\lesssim$10$^{19}$ cm$^{-3}$. The large compensation in the films grown at relatively higher N$_2$ percentages is attributed to the presence of acceptor centers, primarily due to N interstitials, as discussed above. As also mentioned above, N interstitials are substantially reduced in the films grown at lower N$_2$ percentages, having much higher carrier concentrations (~10$^{20}$ cm$^{-3}$), which display significantly reduced values of compensation ratio, as seen from figure 6. The decrease of compensation ratio with decrease of $N_2$ percentage is mainly attributed to the increase in donor concentrations due to Ga interstitials and decrease of N interstitials, as discussed above. It may be mentioned that our earlier XAS studies [39] have also indicated the presence of other acceptor centers, namely, O interstitials, Ga vacancies and possibly $V_{Ga}$-$O_N$ complexes, which may also partly contribute to the compensation in these films. Thus, the increase of mobility with carrier concentration in undoped and Si-doped GaN films is primarily attributed to the decrease of compensation ratio, although the dominance of ionized impurity scattering restricts the overall mobility to moderate values.

In order to further ascertain the transport mechanism in undoped and Si-doped GaN films, the temperature dependence of the electrical resistivity, mobility and carrier concentration was studied in the temperature range of 10 K to 300 K. Figure 7 shows the temperature dependent resistivity, mobility and carrier concentration of typical undoped and Si-doped GaN films grown at 30% or lower $N_2$ in sputtering atmosphere, having carrier concentrations of ~$10^{19}$ cm$^{-3}$ or higher. It may be mentioned that reliable temperature dependent Hall data could not be obtained for the high resistivity films grown at 40% and higher $N_2$, particularly in the case of undoped films. It is seen that the resistivity of both undoped and Si-doped GaN films decreases substantially with increase of temperature, dominated by the increase of mobility, but accompanied in all the cases by a small increase of carrier concentration. It is also noted that the changes in carrier concentration and mobility with temperature become smaller with decrease of $N_2$ percentage and the consequent increase of carrier concentration. It is however found that the observed change in carrier concentration with temperature in all the cases, is too small to account for donor activation in undoped and Si-doped GaN films, which have been reported to be in the range of 7 - 58 meV [30,58] and 10 - 30 meV [34,59-61], respectively. It is also seen from figure 7 that in both undoped and Si-doped GaN films, the change of mobility with temperature is substantially larger than the corresponding change in carrier concentration. It is noteworthy that the temperature dependence of mobility is similar for both undoped and Si-doped GaN films, displaying a nearly linear increase with temperature in the entire range of measurement.

In order to understand the temperature dependence of mobility of undoped and Si-doped GaN films, the mobility data has been analyzed by taking into account the scattering due to charged dislocations and ionized impurities, as was done to analyze the room temperature mobility data. It may be mentioned that the small increase of mobility with temperature could not be fitted using the grain boundary scattering model [61-64], and moreover, the order of magnitude of grain boundary potential/pseudo potential [64,65] estimated near room temperature from the mobility data was found to be insignificant (≲1 meV). As discussed above, the room temperature mobility of the films having carrier concentrations ≲$10^{19}$ cm$^{-3}$ is controlled by contributions to scattering from ionized impurities and charged dislocations, the former being determined by high compensation ratio. On the other hand, in the case of highly degenerate films with carrier concentrations approaching $10^{20}$ cm$^{-3}$, the mobility is primarily determined by ionized impurity scattering. Following a similar methodology as above, the measured mobility was fitted by calculating $\mu_{ii}$ and $\mu_{dis}$ at all temperatures and invoking Matthiessen's rule, with the compensation ratio as a fitting parameter. Figures 8 and 9, respectively show the results of these fittings for undoped and Si-doped GaN films. For undoped GaN films, the calculated values of $\mu_{ii}$ and $\mu_{dis}$, used for fitting the temperature dependent mobility data are plotted in figure 8(a), along with the corresponding values of compensation ratio as a function of temperature in figure 8(b), while the calculated and measured values of mobility are plotted in figure 8(c). Similarly, figure 9(a,b,c) shows the corresponding plots for Si-doped GaN films. In both figures 8(a) and 9(a), $\mu_{dis}$ is found to decrease slightly with decrease of temperature, due the marginal decrease of carrier concentration with temperature (as shown in figure 7), following equation 2. On the other hand, in equation 1, the decrease of carrier concentration with temperature is expected to cause increase of $\mu_{ii}$. However, $\mu_{ii}$ is found to decrease slightly with temperature, because, in order to fit the measured mobility *vs.* temperature data, the calculated values of $\mu_{ii}$ were obtained by increasing the compensation ratio. The corresponding dependences of the compensation ratio on temperature are shown in figures 8(b) and 9(b). Finally, figures 8(c) and 9(c) show that the calculated and measured values of mobility are in good agreement for both undoped and Si-doped GaN films, in the entire temperature range. Thus, the observed mobility behaviour of these films can be explained by the combined effect of scattering due to ionized impurities and charged dislocations, with the former being determined by a temperature dependent compensation ratio.

The small increase of compensation ratio with decrease of temperature seen in both undoped and Si-doped GaN films, can be explained as follows. It is generally known that in degenerate semiconductors, the Fermi level is positioned within the conduction band and hence, the carrier concentration is expected to be temperature independent. However, this is not seen in the present case, wherein, in spite of the high carrier concentration of the films, it is found to decrease marginally with temperature, contributing significantly to the increase of resistivity, as shown in figure 7. Such an increase of resistivity with decrease of temperature has been reported in highly doped semiconductors with substantial compensation and the effect is found to become stronger with increase of compensation [66-68]. This phenomenon has been explained by considering the influence of the Coulomb potential of defects in heavily doped semiconductors [69,70], which lead to fluctuations in the band edges, as described in Refs. [67, 68]. It has been propounded that depending on the position of the Fermi level within the modulated conduction band edge, the electrons fill the 'lakes' within a hilly terrain. At high compensation, the level of the carrier lakes drops, which can cause a substantial decrease of conductivity and the degenerate semiconductor reverts from metallic conduction to thermal activation over saddle points in the hilly terrain [68]. Gadzhiev *et al*. [66] have reported the effect of compensation in highly doped *n*-type Ge (with As concentration of $8 \times 10^{18}$ cm$^{-3}$), which showed metallic behaviour in the uncompensated sample. However, in the case of compensated samples, the resistivity was found to increase with decrease of temperature, displaying an activation energy, which decreased with temperature. According to them, at high temperatures, the activation energy corresponds to the difference between the percolation level and Fermi level, while at low temperatures, hopping conduction dominates [66].

In the present case, both undoped and Si-doped GaN films are degenerate and as discussed above, are substantially compensated due to the presence of acceptor centers. As the temperature dependence of resistivity, shown in figure 7, displays a behaviour similar to that described above, it is proposed that this behaviour arises essentially from the effects due to band edge fluctuations, which are mainly responsible for the decrease of carrier concentration with temperature. However, the effect appears to be relatively smaller in the present case, as it depends on the position of the Fermi level above the fluctuating conduction band edge, which is essentially determined by the extent of doping and compensation in the films. It is noteworthy that the observed effects are relatively prominent in the films with lower carrier concentrations ($\lesssim 10^{19}$ cm$^{-3}$) and high compensation ratio, and become marginal in the films with high carrier concentrations and low compensation ratio. Furthermore, considering that in substantially compensated films, the ionized acceptors ($N_A^-$) may not be significantly affected by the decrease of temperature, the marginal decrease of carrier concentration with temperature can be attributed to the decrease of ionized donors ($N_D^+$) at lower temperatures, which results in the observed increase of compensation ratio.

## 4. Conclusion

The electrical properties of undoped and Si-doped GaN films (with nearly constant Si content of ~2 at. %) grown on sapphire by reactive sputtering are found to be strongly dependent on the N$_2$ percentage in sputtering atmosphere. The resistivity of the Si-doped GaN film grown at 100% N$_2$ is five orders of magnitude smaller than that of the undoped film grown under the same condition, confirming the effect of Si doping. With decrease of N$_2$ from 100% to 75%, the carrier concentration of Si-doped GaN films increases from ~7 × 10$^{18}$ cm$^{-3}$ to ~2 × 10$^{19}$ cm$^{-3}$, due to the decrease of N interstitials, resulting in reduced compensation. As the N$_2$ percentage is decreased to ~20%, the carrier concentration displays a plateau-like behaviour, which is explained by the combined effects due to the saturation of Si doping and increase of Ga interstitial donors, as well as the compensation by acceptors due to N interstitials and Ga vacancies. Interestingly, undoped and Si-doped GaN films grown below 20% N$_2$ display nearly the same values of electrical parameters and similar dependences on N$_2$ percentage, indicating the prevalence of the same source of *n*-type conductivity in both cases, which is attributed primarily to Ga interstitial donors. Both undoped and Si-doped GaN films are degenerate and display a concurrent increase of mobility with carrier concentration. Analysis of the room temperature mobility data shows that the low mobility of undoped and Si-doped GaN films having carrier concentrations ≲10$^{19}$ cm$^{-3}$ is controlled by the contributions from scattering due to ionized impurities and charged edge dislocations, the former being determined by a high compensation ratio (~0.9). With increase of carrier concentration above 10$^{19}$ cm$^{-3}$, the effect of dislocation-controlled scattering becomes insignificant due to stronger screening as well as the decrease in the density of edge dislocations. Accordingly, in the case of highly degenerate films with carrier concentrations of ~10$^{20}$ cm$^{-3}$, the mobility is determined primarily by ionized impurity scattering. The concurrent increase of mobility with carrier concentration in both undoped and Si-doped GaN films is attributed to the large reduction in compensation ratio (to ~0.5) due to the substantial decrease of N interstitials in the films grown at low N$_2$ percentages. The temperature dependent transport measurements reveal that in spite of their degenerate character, both undoped and Si-doped films display a small decrease of carrier concentration along with a nearly linear decrease of mobility, with decrease of temperature. The temperature dependent mobility data has also been analyzed by considering the combined effect of scattering due to ionized impurities and charged dislocations, where the former is influenced by a marginal increase of compensation ratio with decrease of temperature. The increase of compensation ratio is explained by the reduction of ionized donors, in consonance with the observed decrease of carrier concentration with temperature, and is attributed to band edge fluctuation effects in highly doped semiconductors with substantial compensation.

**Acknowledgements**

The Central Surface Analytical Facility, Hall Measurement and HRXRD Central Facilities as well as Sophisticated Analytical Instruments Facility (SAIF) of IIT Bombay are greatly acknowledged for respective measurements.

**Author declarations**

**Conflict of interest**

The authors have no conflicts to disclose.

**Data availability statement**

All data that support the findings of this study are included within the article (and any supplementary files).

**Figures**

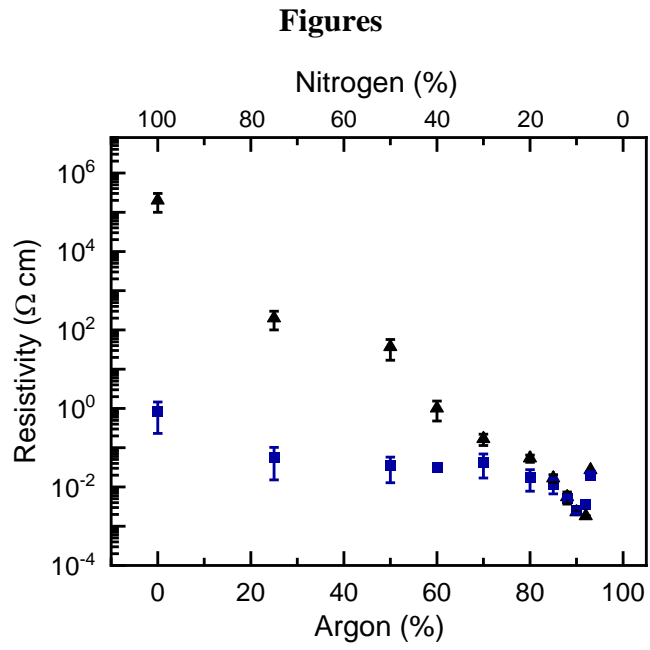

**Figure 1.** Variations of the resistivity of undoped (▲) and Si-doped (■) GaN films with Ar and $N_2$ percentages.

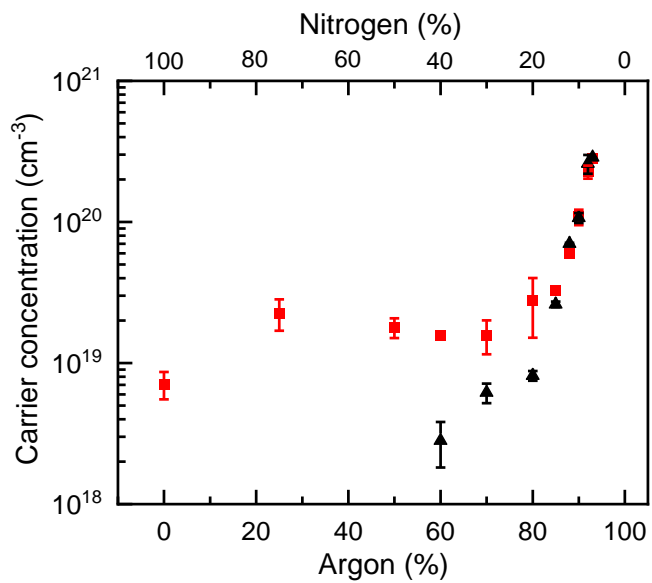

**Figure 2.** Variations of the carrier concentration undoped (▲) and Si-doped (■) GaN films with Ar and $N_2$ percentages.

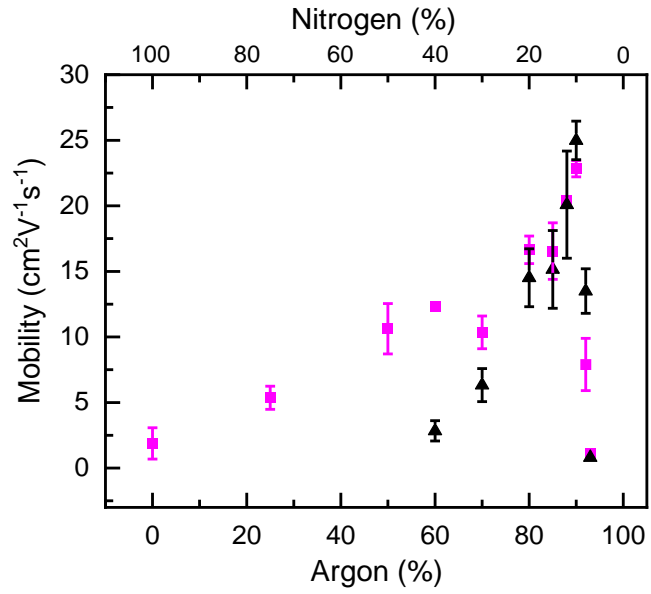

**Figure 3.** Variations of the mobility of undoped (▲) and Si-doped (■) GaN films with Ar and N$_2$ percentages.

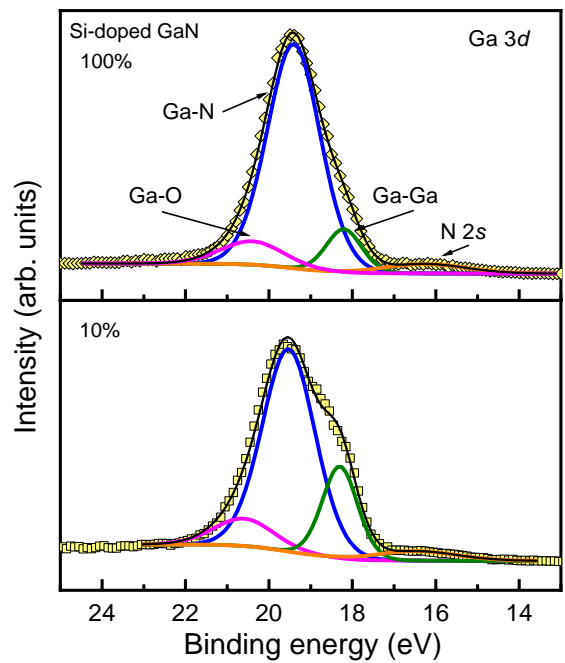

**Figure 4.** Ga 3*d* core level spectra obtained after in-situ surface etching (2 minutes) of Si-doped GaN films grown at 100% and 10% N$_2$ (as indicated).

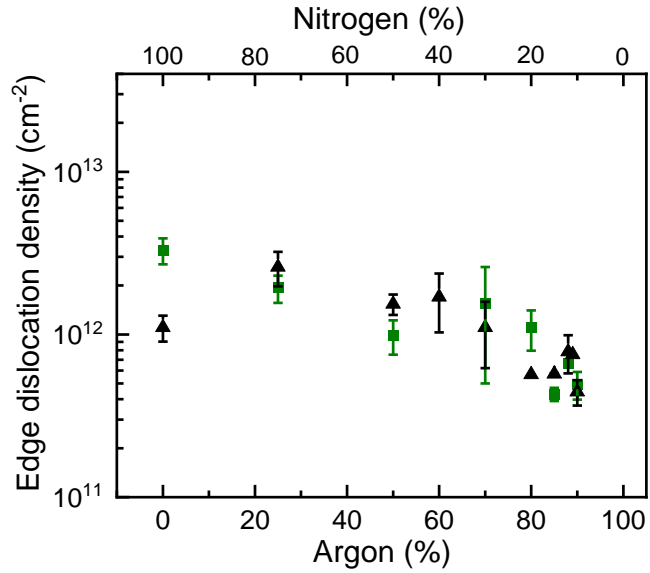

**Figure 5.** Plots of the density of edge dislocations of undoped (▲) and Si-doped (■) GaN films against Ar and $N_2$ percentages.

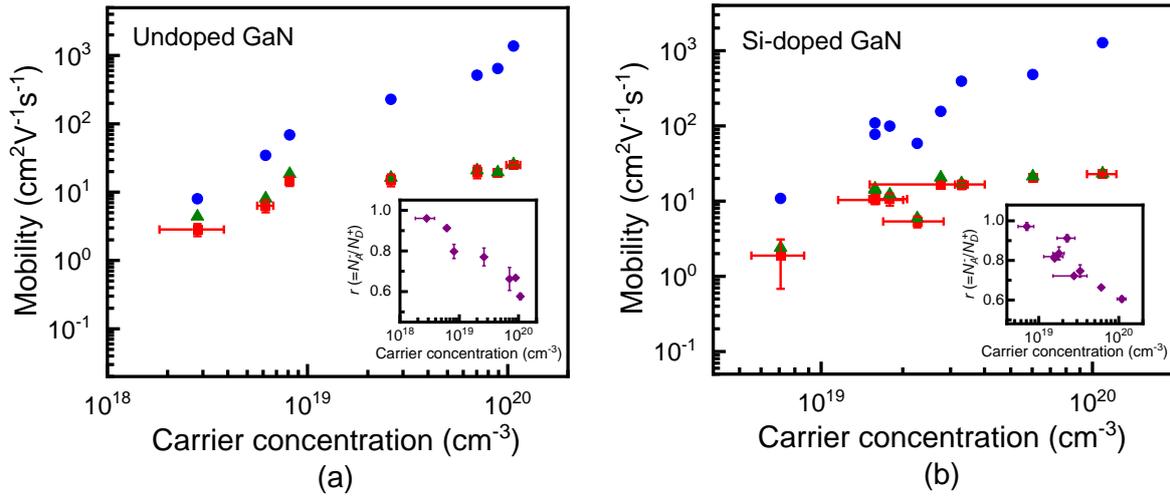

**Figure 6.** Plots of the measured mobility (■) with carrier concentration, along with the calculated values of mobility due to ionized impurity scattering (▲) and dislocation scattering (●) for (a) undoped and (b) Si-doped GaN films. The insets show the corresponding variations of compensation ratio (*r*) with carrier concentration.

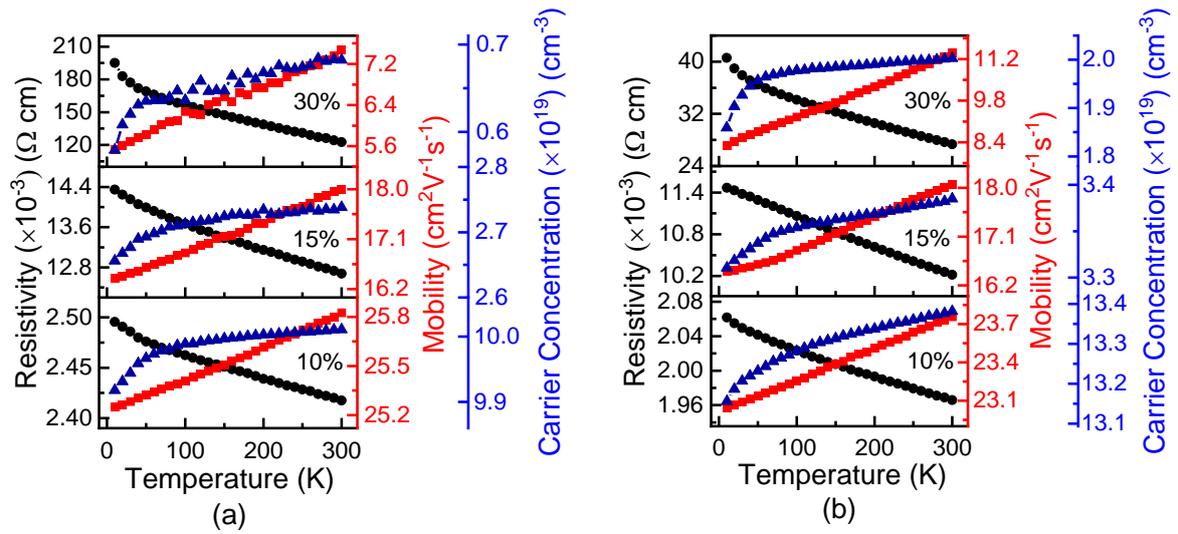

**Figure 7.** Temperature dependence of resistivity (●), mobility (■) and carrier concentration (▲) of (a) undoped and (b) Si-doped GaN films grown at different $N_2$ percentage (as indicated).

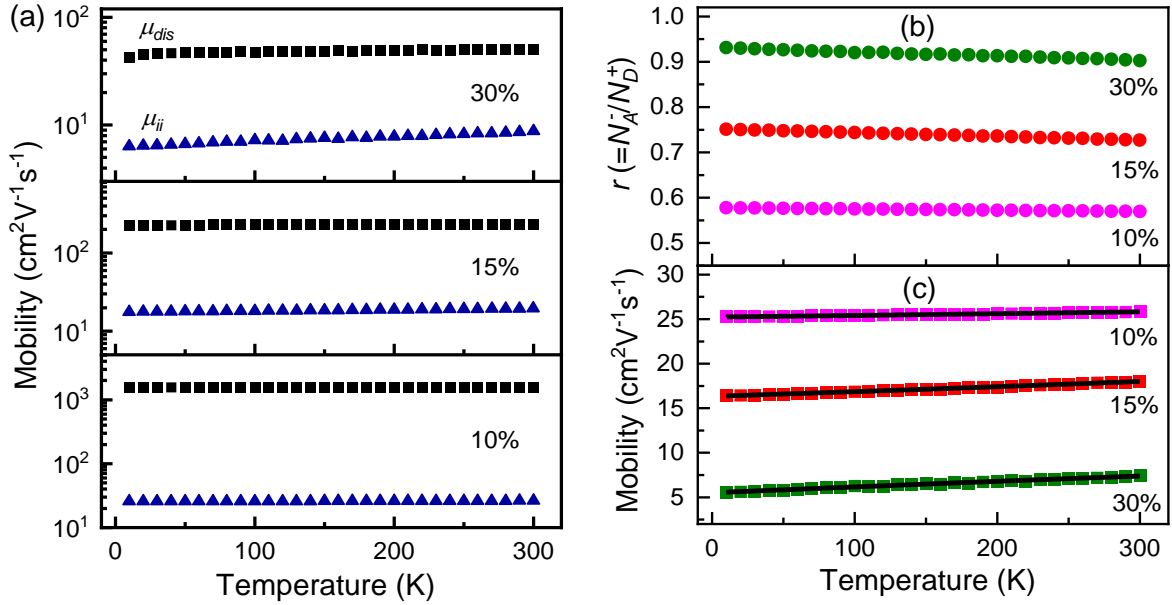

**Figure 8.** The plots of (a) calculated values of μ$_{ii}$ (▲) and μ$_{dis}$ (■), (b) compensation ratio (*r*) and (c) the measured mobility, along with the fitting (solid lines) using ($\mu^{-1} = \mu_{ii}^{-1} + \mu_{dis}^{-1}$), with temperature, for undoped GaN films grown at different N$_2$ percentages in sputtering atmosphere (as indicated).

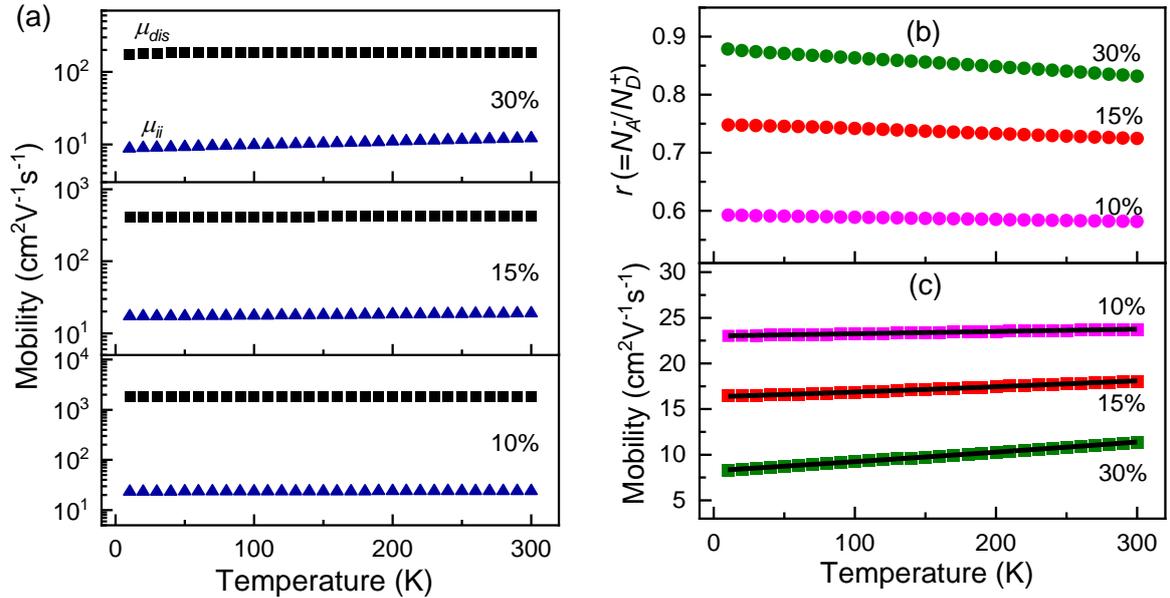

**Figure 9.** The plots of (a) calculated values of μ$_{ii}$ (▲) and μ$_{dis}$ (■), (b) compensation ratio (*r*) and (c) the measured mobility, along with the fitting (solid lines) using ($\mu^{-1} = \mu_{ii}^{-1} + \mu_{dis}^{-1}$), with temperature, for Si-doped GaN films grown at different N$_2$ percentages in sputtering atmosphere (as indicated).